\documentclass[conference]{IEEEtran}
\IEEEoverridecommandlockouts

\usepackage{cite}
\usepackage{amsmath,amssymb,amsfonts}
\usepackage{algorithmic}
\usepackage{graphicx}
\usepackage{textcomp}
\usepackage{xcolor}
\def\BibTeX{{\rm B\kern-.05em{\sc i\kern-.025em b}\kern-.08em
    T\kern-.1667em\lower.7ex\hbox{E}\kern-.125emX}}
\begin{document}

\title{Toward Near-Real-Time Marine Oil Spill Detection in SAR Imagery using Quantum-Assisted SVM\\
}

\author{\IEEEauthorblockN{Joseph Strauss}
\IEEEauthorblockA{\textit{Division of Computer Science \& Engineering} \\
\textit{Louisiana State University}\\
Baton Rouge, United States \\
jstrau9@lsu.edu}
\and
\IEEEauthorblockN{ Dr. Jyotsna Sharma}
\IEEEauthorblockA{\textit{Department of Petroleum Engineeering} \\
\textit{Louisiana State University}\\
Baton Rouge, United States \\
jsharma@lsu.edu}
}

\maketitle

\begin{abstract}
Marine oil spills require rapid detection to mitigate severe ecological and economic damage. While satellite-based Synthetic Aperture Radar (SAR) provides essential all-weather monitoring, analyzing this data remains challenging. Deep learning models often require massive datasets and incur high latency. To address this, a pixel-wise quantum-assisted Support Vector Machine (QSVM) bagging ensemble is developed. Quantum annealing is leveraged to optimize the support vectors of individual weak SVMs on small data subsets, which are then classically aggregated. 

The approach is evaluated on Sentinel-1 imagery using both quantum simulation and physical quantum annealing hardware. The quantum-assisted pipeline achieved performance comparable to a rigorous classical baseline, yielding an Intersection-over-Union (IoU) of 0.60 and a balanced accuracy of 0.89. Complementary experiments with gate-based quantum computing demonstrated similar segmentation accuracy, although the annealing approach offered superior inference efficiency.  Generalization was further assessed on independent oil spill imagery from the Strait of Hormuz, demonstrating the potential transferability of the trained pipeline to geographically distinct spill events. These results establish the feasibility of quantum-assisted, segmentation pipelines for near-real-time environmental monitoring. 
\end{abstract}


\section{Introduction}
Marine oil spills present acute ecological and economic risks, necessitating rapid and accurate detection critical for environmental protection and mitigation efforts. Satellite-based Synthetic Aperture Radar (SAR) has emerged as a promising tool for this task due to its all-weather, 24 hour, large-area monitoring capabilities. However, translating SAR data into actionable intelligence remains challenging, as supervised machine learning approaches require high-quality, pixel-level annotated data and and deep learning models often incur substantial latency and data requirements. In this regime, lightweight models such as Support Vector Machines (SVMs) offer a compelling alternative for low-latency, data-efficient inference.

Quantum machine learning introduces a complementary pathway by enabling efficient exploration of high-dimensional, non-convex optimization landscapes. In particular, quantum annealing–based formulations have shown that weak SVM learners can be optimized as discrete combinatorial problems and grouped into ensembles, potentially yielding improved generalization in low-data regimes. Building on this paradigm, this work extends quantum-enhanced ensemble learning from image-level classification to dense, pixel-wise segmentation for SAR-based oil spill detection.

A pixel-wise classification framework is developed in which multiple weak SVM classifiers are trained on small, randomly sampled subsets of labeled SAR data and combined into a bagging ensemble. Optimization of each classifier is carried out via quantum annealing. This formulation preserves the data efficiency of classical ensembles while leveraging quantum hardware to efficiently navigate the complex optimization landscapes underlying individual model training.

Empirical evaluation on Sentinel-1 SAR imagery, conducted on both quantum simulation and physical quantum annealing hardware, shows that the proposed quantum-assisted segmentation approach achieves performance comparable to a rigorously controlled classical bagging SVM baseline, with Intersection-over-Union (IoU) of 0.60 and balanced accuracy of 0.89. Additional experiments were carried out to assess the performance of gate-based quantum computing on image segmentation tasks. While the gate-based and annealing-based QSVM workflows demonstrate similar segmentation accuracy, the annealing-based approach exhibits more favorable inference characteristics for the evaluated task. Generalization is further assessed using independent oil spill imagery from the Strait of Hormuz, demonstrating the potential transferability of the trained pipeline to geographically distinct spill events.  Collectively, these results establish the feasibility of quantum-assisted SVM ensembles for SAR-based oil spill segmentation and highlight their potential as lightweight components of near-real-time environmental monitoring systems.

\section{Background and Related Work}
Marine oil spills represent a persistent and high-impact environmental challenge, with consequences spanning marine ecosystems, coastal economies, and global energy infrastructure. Rapid and accurate detection is critical for enabling timely containment, mitigation, and environmental response \cite{6235983}. Among remote sensing modalities, Synthetic Aperture Radar (SAR) has emerged as the primary tool for marine oil spill detection due to its all-weather, day-and-night imaging capability \cite{ALPERS2019429}. Oil slicks are detected indirectly through their suppression of short-scale ocean surface roughness, which reduces radar backscatter and produces characteristic dark regions in SAR imagery. However, SAR interpretation remains challenging because look-alike phenomena, including low-wind zones and biogenic films, can generate similar dark signatures. This ambiguity necessitates robust classification frameworks capable of discriminating subtle spatial and statistical patterns under geographically diverse acquisition conditions. 

Machine learning has become central to addressing this challenge by framing oil spill detection as an unbalanced image segmentation problem. Each pixel in a SAR image is assigned a class label, typically “oil” or “non-oil.” Performance is commonly evaluated using metrics that are more robust to class imbalance than overall accuracy, including Intersection-over-Union (IoU), F1-score, and balanced accuracy. With today’s emphasis on parallel computing, deep learning architectures such as U-Net have been shown to effectively classify and then segment images containing oil spills \cite{TRUJILLOACATITLA2024116549}. However, while deep learning frameworks excel at complex segmentation, they typically rely on heavily expanded feature spaces. Classical methods, including Random Forests \cite{rs13112044} and ensembles of Support Vector Machines (SVMs) \cite{CHENG2024117012}, provide strong baselines, particularly in low-data regimes. SVMs are attractive in low-data regimes because they construct maximum-margin decision boundaries in high-dimensional feature spaces, often requiring fewer labeled examples than deep neural networks. Ensemble strategies, such as bagging, further improve robustness by aggregating multiple weak learners trained on subsampled data. However, the computational bottleneck lies in the optimization of the base learners. For each SVM, identifying the optimal support vectors requires solving a complex quadratic programming problem \cite{9323544}.

To achieve optimal segmentation without the computational overhead of deep neural networks, Quantum Machine Learning offers novel optimization paradigms. Gate-based models such as Quanvolutional Neural Networks \cite{henderson2020quanvolutional} have demonstrated that quantum circuits can effectively process and recognize image features.  An alternative paradigm, quantum annealing, is specifically designed to solve discrete optimization problems by mapping them to an energy landscape explored through quantum fluctuations \cite{McGeoch2020TheDA}. In this framework, the training of individual SVM classifiers can be formulated as a quadratic unconstrained binary optimization (QUBO) problem. Quantum annealers, such as those developed by D-Wave Systems, provide hardware implementations of this paradigm, while classical emulators enable controlled experimentation without hardware constraints. The Q-Seg framework \cite{venkatesh2024q} formulates image segmentation as a min-cut problem to be optimized on D-Wave quantum annealing hardware. Quantum annealing has also been applied to ensemble learning techniques. In Cavallaro et al. (2020), quantum annealing is used to optimize an ensemble of weak SVMs for classifying multispectral earth observation data \cite{9323544}. Building upon this foundation, this paper introduces a novel framework that leverages quantum annealing to optimize weak SVMs, grouping them into a robust ensemble specifically tailored for the high-fidelity segmentation of SAR marine oil spill imagery.

In contrast, gate-based quantum computing implements quantum algorithms through sequences of parameterized quantum gates. Quantum SVMs in this paradigm are typically realized via quantum kernel methods or variational quantum circuits \cite{bergholm2018pennylane, havlivcek2019supervised}, which map classical data into high-dimensional Hilbert spaces to improve separability. These approaches offer strong theoretical expressivity, although they require repeated circuit evaluations for training and inference, introducing nontrivial overhead.

A critical consideration in quantum machine learning is the gap between simulated and hardware-executed performance. Emulator-based studies provide controlled conditions that isolate algorithmic behavior without hardware noise and connectivity limitations. In contrast, execution on physical quantum hardware tests the viability of the current technology for real world applications. Understanding this emulator-to-hardware gap is essential for assessing the practical viability of quantum-assisted workflows in real-world applications. Despite growing interest in quantum machine learning, most prior work has focused on image-level classification tasks \cite{9323544}, with limited attention to dense, pixel-wise segmentation problems that are central to remote sensing applications. Furthermore, systematic comparisons across classical methods, annealing-based quantum approaches (both emulated and hardware-executed), and gate-based quantum implementations remain scarce, particularly in low-data, latency-sensitive settings.

This work addresses these gaps by developing a unified pixel-wise classification framework for SAR-based oil spill detection and evaluating four configurations: a classical SVM ensemble baseline, a quantum annealing-based SVM implemented on both emulator and hardware, and a gate-based quantum SVM evaluated via circuit simulation. This structured comparison enables isolation of algorithmic and hardware effects, providing insight into performance, generalization, and computational trade-offs across quantum paradigms in operationally relevant remote sensing scenarios.

\section{Methodology}
The proposed quantum-assisted SVM framework was evaluated using Sentinel-1 SAR oil spill imagery and corresponding ground-truth segmentation masks compiled from Trujillo-Acatitla et al. (2024). The dataset consisted of 400 training images and 30 validation images. Each SAR image was downscaled to 256x256 pixels prior to feature extraction and model training.  Based on the validation set of 30 such images, several preprocessing operations were applied before training and testing the model to improve image quality and reduce confounding effects before training and inference. These steps include median blurring, clipping extreme values, masking out land, and performing gamma correction based on NIQE score \cite{CHENG2024117012}. 

To construct a balanced training set, a maximum of 50 pixels was sampled from each training image, consisting of up to 25 oil pixels and 25 water pixels when both classes were available. This sampling strategy mitigated the strong class imbalance inherent in SAR oil spill imagery, where oil pixels typically represent only a small fraction of the full scene. In addition, hard negative mining was applied to the water class by preferentially selecting a subset of the darkest water pixels. This step was designed to expose the classifier to challenging non-oil pixels that visually resemble oil slicks in SAR imagery. 

For each sampled pixel, a five-dimensional feature vector was computed using both VV and VH polarization channels. The features were selected to capture local intensity, polarization contrast, texture, and edge information. The first feature was the raw VV-band intensity. The second feature was the VH-to-VV intensity ratio. The third feature was the local entropy of the VV band computed using a 3x3 window. The fourth feature was the local standard deviation of the VV band, also computed using a 3x3 window. The fifth feature was the gradient magnitude, obtained using Sobel filters along the horizontal and vertical image axes. Together, these features provide a compact representation of pixel-level SAR backscatter behavior and local spatial variability.  

After feature extraction, a quantum-assisted bagging SVM ensemble was trained using an architecture adapted from \cite{9323544}. Each base learner was trained on a disjoint subset of 40 feature vectors. A total of 500 base learners were constructed so that the sampled feature space was broadly represented across the ensemble. Unlike prior applications focused on image-level classification, the proposed framework performs classification at the pixel level. During inference, each pixel in a SAR image is assigned an oil or water label, and the ensemble predictions are aggregated to generate a binary segmentation mask representing the predicted oil spill extent.

For the annealing-based implementation, the optimization of each weak SVM learner was formulated as a quadratic unconstrained binary optimization (QUBO) problem. The continuous SVM coefficients were discretized into binary variables, allowing the SVM training objective to be mapped onto an annealing-compatible optimization problem. Each QUBO instance was then solved using either a D-Wave quantum annealer or a quantum annealing simulator. The resulting binary solutions were decoded to obtain the SVM coefficients and bias term for each weak learner. During inference, each trained weak learner evaluated the feature vector of a candidate pixel, and the final ensemble prediction was obtained by aggregating the outputs across the 500 classifiers. 

To compare the annealing-based approach with a gate-based quantum implementation, a quantum kernel SVM ensemble was also implemented using PennyLane \cite{bergholm2018pennylane}.  A five-qubit quantum circuit was used, with one qubit corresponding to each input feature. The feature vector was encoded through parameterized rotation gates. For each pair of samples, the first feature vector was encoded using positive rotations and the second using negative rotations, allowing the circuit measurement outcomes to quantify sample similarity in the quantum feature space. These pairwise similarities were used to construct a quantum kernel matrix, which was then supplied to a classical SVM optimizer implemented in Scikit-Learn \cite{JMLR:v12:pedregosa11a}. 

This gate-based approach differs from the annealing-based implementation in where the quantum computation enters the workflow. In the annealing approach, quantum computation is used during training to optimize the weak SVM learners, while inference is performed classically using the learned coefficients. In the gate-based approach, SVM optimization is performed classically, but quantum circuits are evaluated during kernel construction and inference. Consequently, each test pixel must be encoded into the five-qubit circuit and compared against the support vectors before the classical decision function can be evaluated. This distinction enables a direct comparison of the computational trade-offs between annealing-based training acceleration and gate-based quantum-assisted inference for pixel-wise SAR oil spill segmentation.

\section{Results}
Experiments were conducted on a subset of a publicly available Sentinel-1 SAR oil spill dataset with corresponding ground-truth segmentation masks \cite{TRUJILLOACATITLA2024116549}. The test set contained 150 unseen images. Classical computations were performed on a workstation equipped with a 12-core AMD Ryzen 3900X CPU and an NVIDIA RTX 3080 GPU. Quantum annealing experiments were executed on D-Wave’s Advantage\_system4.1 quantum annealer. 

All quantum annealing workflows were implemented using D-Wave’s Ocean SDK, which was used for QUBO construction, minor embedding, sampler configuration, and annealing simulation. For hardware execution, each QUBO problem was submitted with 1000 reads to account for the stochastic nature of quantum annealing and the possibility that individual samples may converge to near-optimal rather than globally optimal solutions. The best 20 returned samples were retained for decoding and classifier construction. 

The evaluation was organized around four model configurations: a classical bagging SVM baseline, a simulated annealing-based SVM ensemble, a hardware-executed quantum annealing SVM ensemble implemented on D-Wave, and a simulated gate-based quantum SVM ensemble implemented in PennyLane. This structure enables comparison across classical, annealing-based, hardware-executed, and gate-based quantum-assisted workflows under a unified pixel-wise segmentation framework. Quantitative results are summarized in Table~\ref{tab:placeholder}. Representative qualitative segmentation outputs are shown in Fig. 1, including examples of high-, moderate-, and low-quality predictions to illustrate the range of model performance across test images.

\begin{table*}[!t]
    \centering
    \caption{Test Performance Metrics of SVM Models}
    \label{tab:placeholder}
    \begin{tabular}{cccccc}
        Model & IoU & F-1 Score & Balanced Accuracy & Inference time per image (s) & Training time (s) \\
        \hline
        Classical SVM & 0.60 & 0.73 & 0.89 & 1.04 & 13.24 \\
        QSVM - simulated annealing & 0.60 & 0.72 & 0.89 & 2.62 & 113.98 \\
        QSVM - hardware annealing & 0.60 & 0.72 & 0.89 & 2.61 & 8.00 \\
        QSVM - simulated gate model & 0.60 & 0.72 & 0.89 & 23.78 & 3.77 \\
    \end{tabular}
\end{table*}

\begin{figure*}[t]
    \centering
    \includegraphics[width=\textwidth]{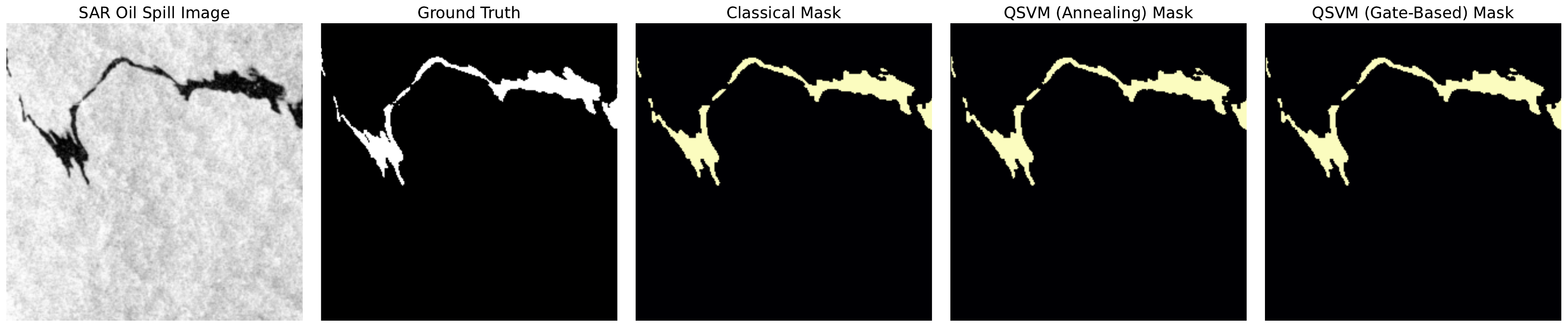}
    \vspace{2mm} 
    
    \includegraphics[width=\textwidth]{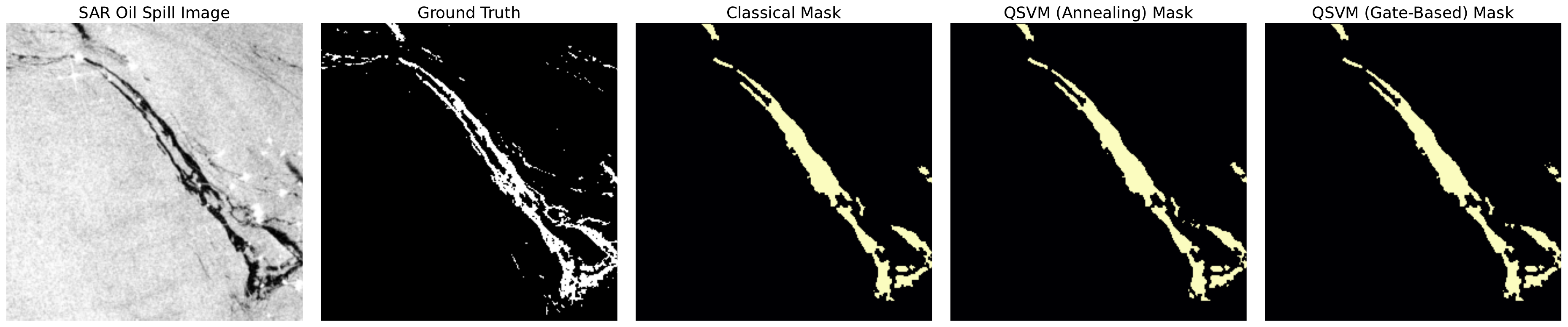}
    \vspace{2mm}
    
    \includegraphics[width=\textwidth]{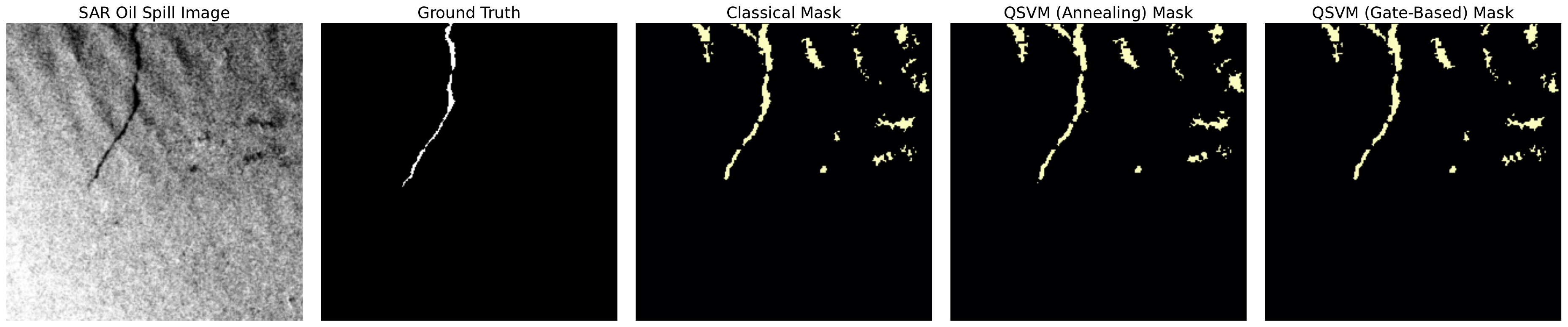}
    
    \caption{Qualitative comparison of segmentation performance across classical and quantum architectures. Top row: High-fidelity segmentation on an easily separable spill. Middle row: Intermediate performance exhibiting partial boundary degradation. Bottom row: Poor segmentation demonstrating susceptibility to complex speckle noise and look-alikes. Columns from left to right: SAR Image, Ground Truth Mask, Classical SVM prediction, QSVM (Annealing) prediction, and QSVM (Gate-Based) prediction.}
    \label{fig:qualitative_comparison}
\end{figure*}

The empirical results show broadly comparable segmentation performance across the tested architectures. This consistency indicates that both the QUBO formulation used for the D-Wave annealing workflow and the quantum-kernel formulation used in the gate-based model preserve the relevant decision structure of the classical SVM ensemble without substantially degrading predictive performance. Thus, under the evaluated conditions, the main distinction among the models is not segmentation quality but computational behavior, particularly mean inference time.  

The classical bagging SVM baseline achieved the fastest inference, requiring 1.04 seconds per image. The simulated and hardware-executed annealing-based approaches remained competitive, with mean inference times of approximately 2.62 seconds per image. This efficiency is expected because the annealing-based models use quantum computation during training to optimize the weak learners, while inference is performed classically using the learned SVM coefficients and can therefore benefit from conventional CPU/GPU acceleration. In contrast, the gate-based quantum SVM exhibited the slowest inference, requiring 23.78 seconds per image. This increased cost arises because the quantum kernel must be evaluated during inference, requiring each test pixel to be encoded into the quantum circuit and compared against support vectors. These results suggest that, among the quantum-assisted approaches evaluated, the annealing-based SVM ensemble provides the most practical balance between segmentation performance and inference efficiency for SAR oil spill segmentation. Fig.~\ref{fig:qualitative_comparison} compares the SAR input images, corresponding ground-truth masks, and predicted segmentation outputs for representative test cases. 

\begin{figure*}[!htbp]
    \centering
    \includegraphics[width=\textwidth]{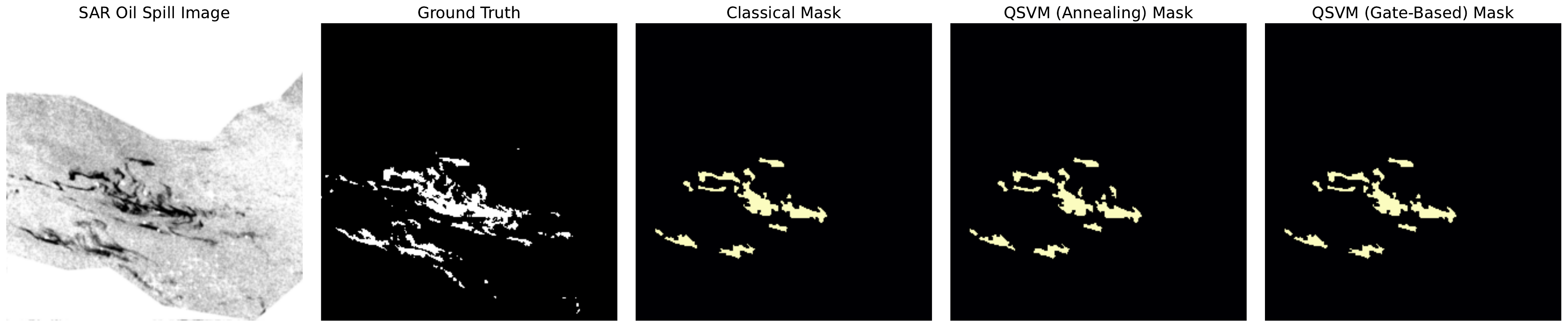}
    \caption{Strait of Hormuz oil spill segmentation results. Columns from left to right: SAR Image, Ground Truth Mask, Classical SVM prediction, QSVM (Annealing) prediction, and QSVM (Gate-Based) prediction.}
    \label{fig}
\end{figure*}

To further assess generalizability, the trained pipeline was evaluated on independent Sentinel-1 SAR imagery of an oil spill in the Strait of Hormuz associated with recent regional conflict. The models were applied to this geographically distinct scene without fine-tuning or recalibration. The results again show comparable performance across the classical and quantum-assisted configurations. The quantum annealing-based model achieved an IoU of 0.46 and a balanced accuracy of 0.75. The gate-based quantum approach achieved an IoU of 0.45 and a balanced accuracy of 0.74, while the classical bagging SVM baseline achieved an IoU of 0.44 and a balanced accuracy of 0.73. Qualitative segmentation results for this independent Strait of Hormuz case are shown in Fig.~\ref{fig}, illustrating the corresponding SAR input, ground-truth mask, and model-predicted oil spill extents. 

Although these metrics are lower than those obtained on the original test set, the decrease is expected given differences in acquisition conditions, scene characteristics, and preprocessing relative to the dataset compiled in Trujillo-Acatitla et al. (2024). Importantly, the models retained the ability to identify the dominant spatial extent and boundary structure of the oil slick in an unseen geographic setting. As shown in Figs. ~\ref{fig:qualitative_comparison} and ~\ref{fig}, the quantum-assisted models produce segmentation masks that are qualitatively consistent with the classical bagging SVM baseline, supporting the robustness of the overall pixel-wise ensemble framework. These findings suggest that the proposed quantum-assisted SVM approach has potential transferability beyond the original dataset, while also underscoring the need for broader validation across additional spill events, sea states, imaging geometries, sensor conditions, and preprocessing pipelines.

\section{Conclusion}
This work demonstrates the feasibility of quantum-assisted machine learning for SAR-based marine oil spill detection. Among the approaches evaluated, the annealing-optimized SVM ensemble provides high-quality pixel-wise segmentation while preserving efficient classical inference suitable for near-real-time monitoring. Future work will focus on improving robustness to SAR-specific noise, oil-spill lookalikes, and other imaging anomalies, while developing additional quantum-assisted machine learning formulations for image segmentation. 

\vfill\eject

\bibliographystyle{IEEEtran}
\bibliography{references}

\vspace{12pt}

\end{document}